\documentclass[12pt,a4paper]{article}
\usepackage[utf8]{inputenc}
\usepackage{amsmath}
\usepackage{amsfonts}
\usepackage{amssymb}
\usepackage{graphicx}
\usepackage[left=2cm,right=2cm,top=2cm,bottom=2cm]{geometry}

\begin{document}

\setcounter{page}{0}
\newpage
\thispagestyle{empty}


\newpage
\begin{center}
\begin{large}
\textbf{Statistics of Projected Motion in one dimension of a d-dimensional Random Walker}
\end{large}\end{center}
\vspace{1.5cm}
\begin{center}\emph{Jayeeta $Chattopadhyay^{1}$ and Muktish $Acharyya^{2}$ }\\
\vspace{.4cm} 
\emph{Department of Physics, Presidency University,\\
86/1 College Street, Calcutta-700073, INDIA}\\
\vspace{.4cm}                                             
E-mail:$ ^1 $ meghbalikajayeeta@gmail.com \\
E-mail:$ ^2 $ muktish.physics@presiuniv.ac.in\\
         
\end{center}
\vspace{1.5cm}
\textbf{Abstract:}We are studying the motion of a random walker in generalised d-dimensional continuum with unit step length (up to 10 dimensions) and its projected one dimensional motion numerically. The motion of a random walker in lattice or continuum is well studied in statistical physics but what will be the statistics of projected one dimensional motion of higher dimensional random walker is yet to be explored. Here in this paper, addressing this particular type of problem, we have showed that the projected motion is diffusive irrespective of any dimension, however, the diffusion rate  is changing inversely with dimensions. As a consequence, it can be predicted that for the one dimensional projected motion of infinite dimensional random walk, the diffusion rate will be zero. This is an interesting result, at least pedagogically, which implies that though in infinite dimensions there is a diffusion but its one dimensional projection is motionless. At the end of the discussion we are able to make a good comparison between  projected one dimensional motion of generalised d-dimensional random walk with unit step length and pure one dimensional random walk with random step length  varying uniformly between $-h$ to $h$ where $h$ is a `step length renormalizing factor'.
\vspace{2.5 cm}


\newpage
\section{Introduction}
\paragraph*{}The basis of random walk theory can be described by the irregular motion of individual pollen particle, famously studied by botanist Brown(1828), now known as Brownian motion. It is somewhat quite surprising that it was only at the beginning of the twentieth century that random walk was described in the literature but its application is widely spread all over in nature starting from movement of animals, micro-organisms \cite{Ran}, cell migration \cite{Ran}, motion of
reagent molecules in a solution \cite{sto}, process of polymerization \cite{S.M} \cite{Hsu}to today's stock prices\cite{redner}. The absorbing phase transition in a conserved lattice gas with random neighbor particle hopping\cite{lub}, quenched averages for self avoiding walks \cite{dhar} on random lattice is also studied. In biology and
real life random walk problem has a great importance. Apart from opening and closing of cell membrane channels \cite{cell} it has also a great importance in cell migration leading blood vessel growth(angeogenesis) \cite{Ran}. Very recently the first passage property \cite{redner} of a random walker has great importance in various aspects. As for example fluorescence quenching \cite{redner}, in which light emission by a fluorescent molecule stops when it reacts with a quencher; integrate-and-fire neurons \cite{redner}, in which a neuron fires only when a fluctuating voltage level first reaches a specified level; and the execution of buy/sell orders when a stock price \cite{redner} first reaches a threshold. The statistics of Pearson walker in two dimensions for shrinking step size and the transition of the endpoint distribution by varying the initial step size is studied in \cite{redpaper}. Studying random walk in lattice(or continuum) with constant step length is an age old problem. But how does this motion get affected if the walker becomes tired gradually is mentioned in \cite{MA}. Random walk for randomly varying step length and its return probability has been studied recently \cite{AB}
\paragraph*{}
Now a days random walk in higher dimensions is a challenging problem. Study of planer motion of a random walker was started with Pearson\cite{pearson}. Rayleigh was apparently the first to have been study random walk in 3 dimensions(known as random flight)\cite{rayleigh}. Then random walk in higher dimensions has been studied by G.N.Watson \cite{watson}. Arithmetic properties of short uniform random walk in arbitrary dimensions with probability densities in the case of up to five steps has been studied in \cite{1508}. Diffusion limits of the Random walk in higher dimensions by Metropolis algorithm has already been calculated\cite{diffusionlimit}. In the classic paper in 1921, Poyla proved that a simple symmetric random walk on $Z^{d}$ is recurrent for $d \leq 2$
and transient otherwise \cite{poyla}. Recurrence relation for arbitrary dimensions has been studied recently \cite{1601}. As an application of all these, multidimensional random walk has been studied in polymer with an emphasis on natural renormalised renewal structure \cite{polymer} and in probabilistic Road maps for path planning in higher dimensional configuration spaces\cite{robot}

\paragraph*{}  

The study of random walk in higher dimensional continuum and its projected one dimensional motion is also an interesting field of research. In this regard it is to be mentioned that, though in reality we are mainly concerned about 1, 2 and 3 dimensions (D) but the problem is extended to higher dimensions for pedagogical interest and to gain a generalised concept about the statistics of random walk problem irrespective of any dimension. In the paper \cite{PRW}  Boguna,   Porra, Masoliver have generalised the idea of persistent random walk in d-Dimensions and showed how the telegrapher equation is modifying. They also discussed that projected one dimensional motion of higher dimensional persistent random walk may open a new way to address the problem of light propagation through thin slabs. But statistical properties of simple random walk in continuum irrespective of any dimension and its projected one dimensional motion are yet to be explored. In this context it is to be mentioned that according to our knowledge in standard textbooks of statistical mechanics one, two and three dimensional random walk are exactly solvable. So we address that the study of projected motion may open a new way to solve the random walk problem analytically irrespective of any dimension. 
Here, in this paper addressing this particular problem we have tried to find out the following answers computationally.\paragraph{} (1) What is the characteristics of projected one dimensional motion of generalised d-dimensional random walk with unit step length? \paragraph{}(2)  Is it diffusive? \paragraph{} (3) And finally, can we construct a pure one dimensional random walk with which projected motion of generalised d-dimensional random walk can be compared? \paragraph{}. The point(3) is important in the context that if we can get equivalent statistical properties of projected motion from pure one dimensional random walk which is exactly solvable then the exact solution of higher dimensional random walk may be possible.

 In this study by pure 1 dimensional random walk we wanted to say random walk in 1 dimensional continuum and projected random walk of higher dimensions means 1 dimensional projection of higher dimensional random walk. The numerical results of detailed statistical analysis of random walk in dimensions 1, 2, 3, 4, 5, 6, 7, 8, 9, 10  and its projected motion  are reported here. The manuscript is organised as follows: In the next section (section (2)) the model and numerical results observed from the numerical simulation are reported and the paper ends with a concluding remarks mentioned in section (3).  
\section{Model and Results}
 \paragraph{}For simple random walk in 2 dimensional regular lattice a random walker can move in any one of the four directions (up,down,right,left) with equal probability. But if we consider random walker moving in a two dimensional continuum, the walker has infinitely possible directions to choose with equal probability and equal step length. Here in this paper, we have studied random walk in generalised d-dimensional continuum with unit step length and its projected one dimensional motion. Here, we have modeled d-dimensional random walk in such a way so that the magnitude of step length in the corresponding dimensions is unity irrespective of any direction and it is constant of motion. If the step length is unity , then the maximum value of the magnitude of projected step length in any co-ordinate is 1 and the minimum value is 0 and every co-ordinate has only two possible directions which are equally probable to choose (we have used here Cartesian co-ordinate system, as for example in 3 Dimensions, the co-ordinates are $x_{1}, x_{2}, x_{3}$) So including both magnitude and direction the projected step length in every co-ordinate varies between -1 to +1. The  formalism of random walk in generalised d-dimensional continuum is as follows:\\

Suppose we are considering random walk in d-dimensional continuum with unit step length . Let, the projected step lengths in each coordinate are ($r_{1}\prime,r_{2}\prime,r_{3}\prime,.......,r_{d}\prime$). Then the step length in d-dimensions is:
\begin{equation}
R^{\prime}=\sqrt{(r_{1}\prime)^2+(r_{2}\prime)^2+......+(r_{d}\prime)^2}
\end{equation}
 But this not normalised, so the actual projected step length in every co-ordinate so that the step length in d-Dimensions is unity are: 
  \begin{equation}
 r_{1}=\dfrac{r_{1}\prime}{R^{\prime}} ,r_{2}=\dfrac{r_{2}\prime}{R^{\prime}}, r_{3}=\dfrac{r_{3}\prime}{R^{\prime}}......r_{d}=\dfrac{r_{d}\prime}{R^{\prime}}
 \end{equation}

 Now the  step length in d-dimensions becomes 
 \begin{equation}
   R={\sqrt{r_{1}^2+r_{2}^2+......+r_{d}^2}} =1
 \end{equation}
The mathematical rule for a random walker moving in generalised d-dimensional continuum can be expressed as:
 \begin{center}
 $x_{1}(t+1)=x_{1}(t)+r_{1}$\\
 $x_{2}(t+1)=x_{2}(t)+r_{2}$\\
 $x_{3}(t+1)=x_{3}(t)+r_{3}$\\
 .                   .         .\\
 .            .         .\\.            .         .\\.            .         .\\.            .         .\\
\end{center}
\begin{equation}
x_{d}(t+1)=x_{d}(t)+r_{d}
\end{equation}

 Where ($ x_{1}(t), x_{2}(t),.... x_{d}(t)$) are the positions of the random walker at time t and ($ x_{1}(t+1), x_{2}(t+1),.... x_{d}(t+1)$ )are the the positions of the random walker in the next time step, ($ r_{1}, r_{2}, r_{3}......r_{d} $ ) are the projected step lengths in corresponding coordinates varying between -1 to 1 so that the step length in corresponding dimensions is unity.\\
 We have classified this method as \textit{Vector method for generalized d-dimensional random walk} which can be well understood by selecting some points randomly on the surface of a d-dimensional hypersphere of unit radius.

\begin{figure}[hbtp]
\centering
\includegraphics[scale=0.6]{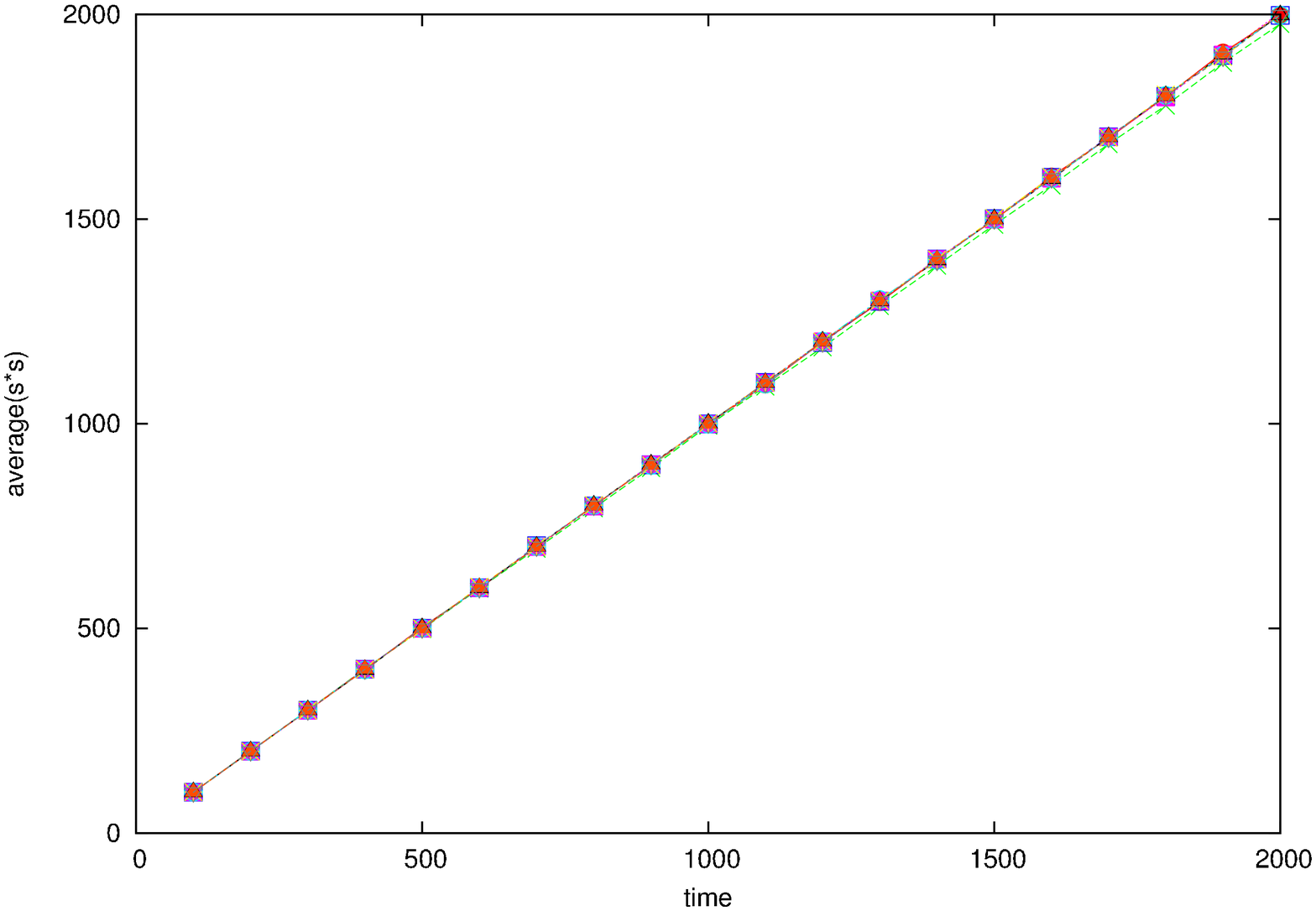}
\caption{Mean-square-displacement ($\langle S^{2}\rangle$) versus time ($ t $) is plotted for 1, 2, 3, 4, 5, 6, 7, 8, 9, 10 dimensions . Here number of samples  $N _{s} $=100000 }
\label{diffusion}
\end{figure}

\begin{figure}[hbtp]
 \centering
 \includegraphics[scale=0.6]{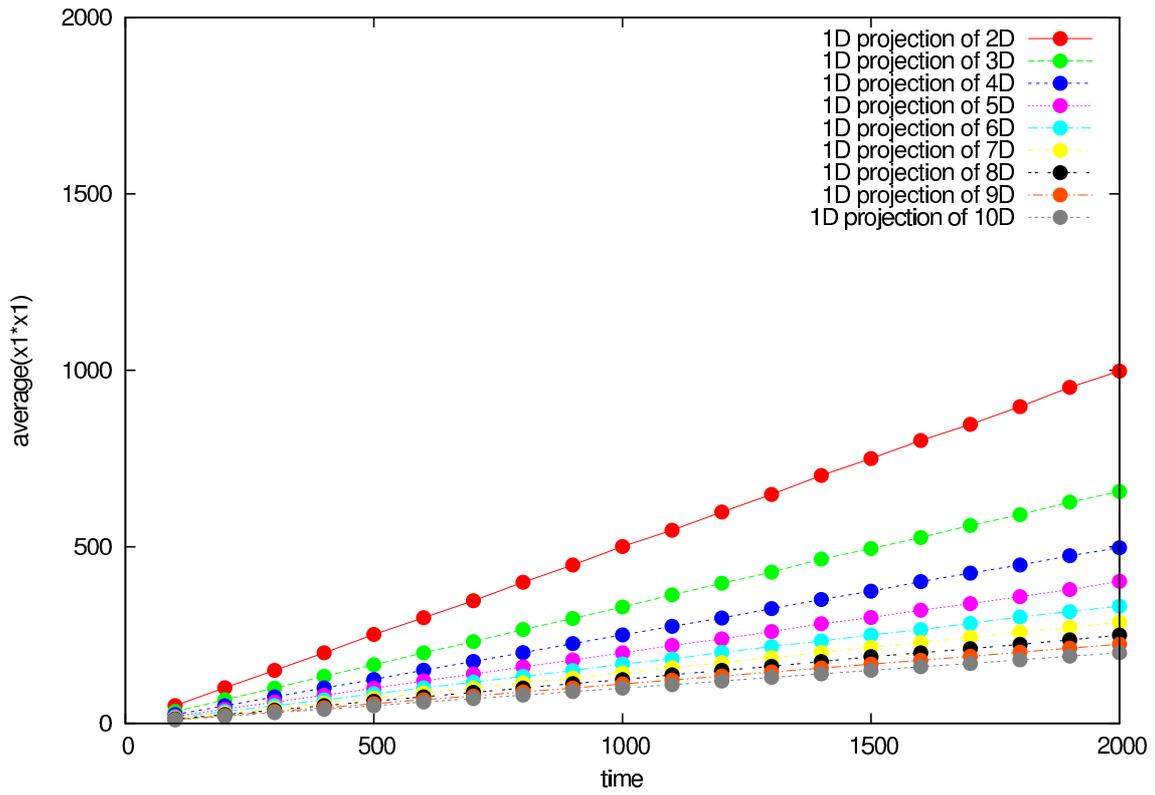}
 \caption{Mean-square-displacement ($\langle x_{1}^{2}\rangle$) of  1 dimensional projection of  d-dimensional random walk (d= 2, 3, 4, 5, 6, 7, 8, 9, 10 dimensions ) versus time $t $ for $N _{s} $=100000. Here it is seen that as we increase the dimension, the slope of the graph i.e diffusion rate decreases.}
 \label{diffnd1d}
 \end{figure}

\begin{figure}[hbtp]
\centering
\includegraphics[scale=0.6]{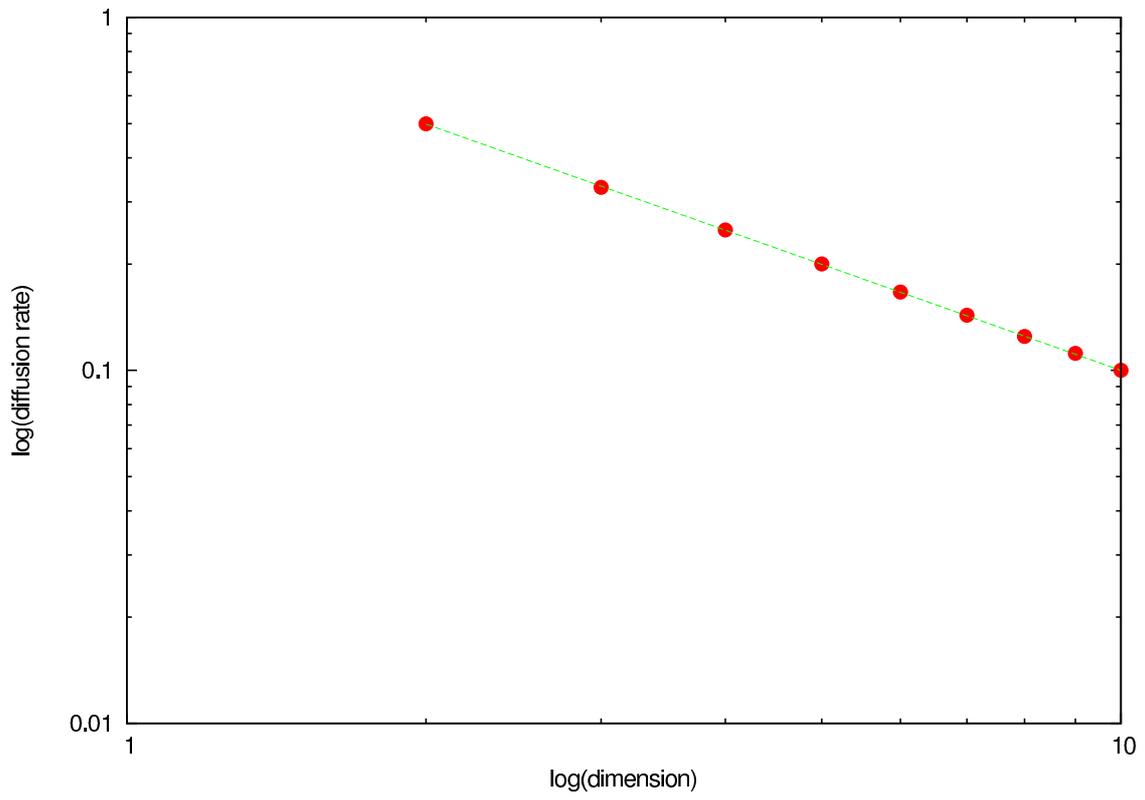}
\caption{ Diffusion rate ($c$) of 1 dimensional projection of  d-dimensional random walk (d= 2, 3, 4, 5, 6, 7, 8, 9, 10 dimensions ) versus dimensions ($d$). The dotted line is $f(x)=1/x$}
\label{diffrate}
\end{figure}

\begin{figure}[hbtp]
\centering
\includegraphics[scale=0.6]{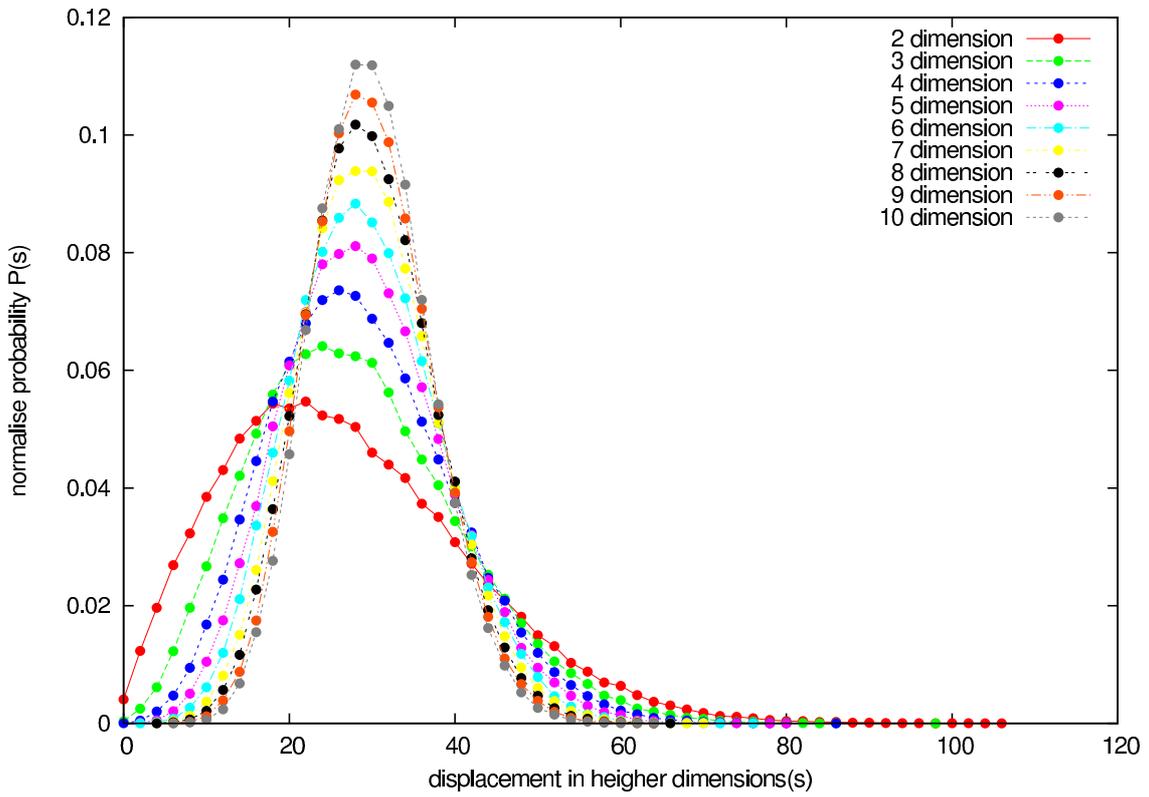}
\caption{Distribution of absolute displacement P(s) after 1000 time steps($ N_{t} $=1000) for  d-dimensional (d= 2, 3, 4, 5, 6, 7, 8, 9, 10 dimensions ) random walk with unit step length.here $ N_{s} $=100000,}
\label{nddists}
\end{figure}

\begin{figure}[hbtp]
\centering
\includegraphics[scale=0.6]{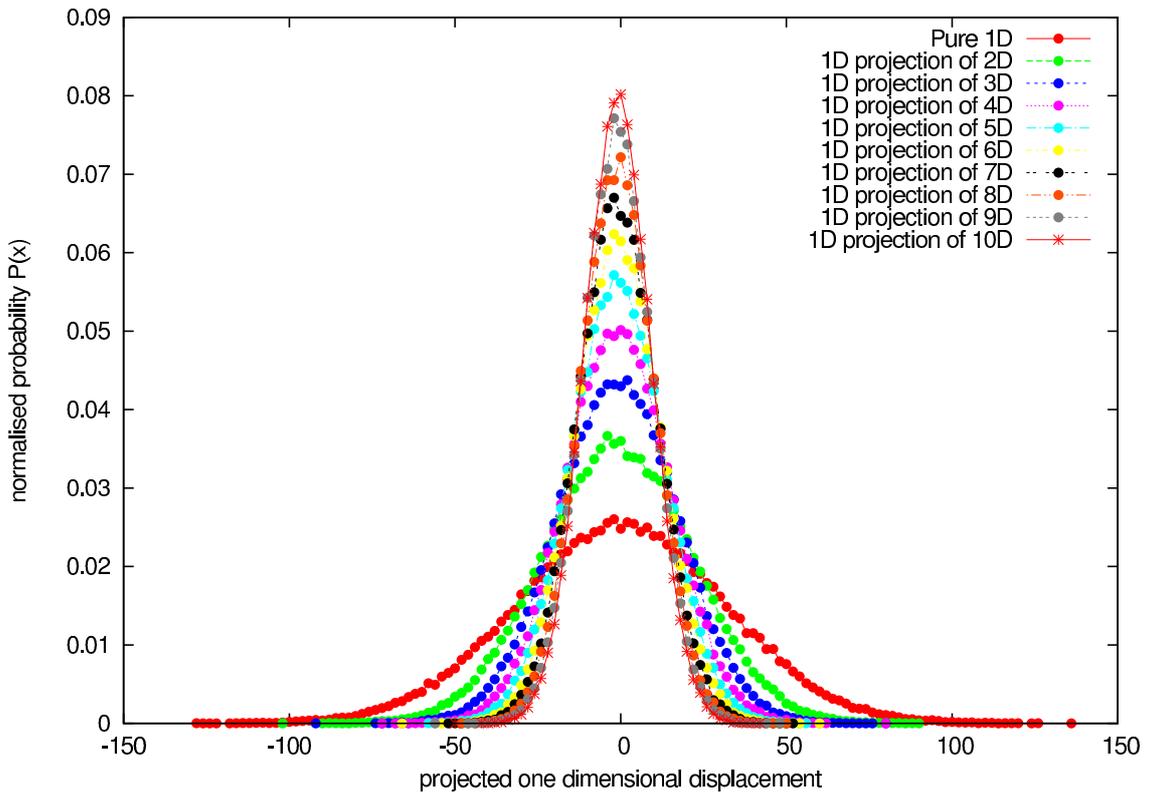}
\caption{Distribution of displacement of 1 dimensional projection of d-dimensional random walk (d= 2, 3, 4, 5, 6, 7, 8, 9, 10 dimensions ) with unit step length and Distribution of displacement for pure one dimensional  random walk with unit step length after the time step $ N_{t} $=1000 for $ N_{s} $=100000. Here it is seen that as dimensions are increasing the distributions are getting sharper and sharper.}
\label{nd1ddistx}
\end{figure}

 \paragraph*{}
 Now we know random walk is purely diffusive. But still for verification we have calculated mean square displacement $ \langle S^{2}\rangle $ for the dimensions (1, 2, 3, 4, 5, 6, 7, 8, 9, 10) which is proportional to time (t) that reveals diffusive behavior (fig(\ref{diffusion})). The distribution of absolute displacement after time steps ($N_{t}=1000$) in d-dimensional continuum  is plotted in (fig\ref{nddists}) where it is seen that the distributions are non-monotonic and unimodal and as we increase the dimension, the most probable value i.e maximum probability of finding the walker at a distance ($S_{m}$) from the starting point is shifting towards higher value.
\paragraph*{}Now the question is what are the properties of the projected motion of generalised d-dimensional random walk. So here we have studied the distribution of displacement ( after $ N_{t} $=1000 steps) of one dimensional projection of higher dimensional random walk (2 Dimensions to 10 Dimensions) (in fig(\ref{nd1ddistx})). From this figure it is observed that all the distributions for projected motion are Gaussian with zero mean but the widths are getting sharper and sharper as we increase the dimensions. From fig(\ref{diffnd1d}) it is observed that projected mean square displacement $ \langle x^{2}\rangle $ of higher dimensions (2, 3,...,10) is proportional to time that reveals pure diffusive nature but the slope of the graphs i.e diffusion rates ($ c=d\langle x^{2}\rangle/dt $ ) are different for different dimensions. So we have plotted $ c $ versus dimension which shows that $ c $ varies inversely with dimension(d).(fig\ref{diffrate}) 
  \begin{equation}
 c = \dfrac{1}{d}
 \label{diff}  
\end{equation}  
   
   From this dependence of $c$ on $d$ it can be predicted that\emph{ in case of infinite dimensions $c$ goes to `0' .This is an interesting result at least pedagogically which can be well interpreted as, though there is a motion in infinite dimensions, but its one dimensional projection is motionless.}.\begin{center}
In the following table numerical values (observed from simulation) of Diffusion rate of projected motion corresponding to dimensions ( 2 to 10 dimension) are mentioned:
\end{center}
\begin{center}
\begin{tabular}{|c|c|}
\hline 
Dimension &  Diffusion rate of projected motion \\ 
\hline 

2 & 0.499 \\
3 & 0.330 \\
4 & 0.250 \\
5 & 0.200 \\
6 & 0.167 \\
7 & 0.143 \\
8 & 0.125 \\
9 & 0.112 \\
10 & 0.100 \\
\hline 
\end{tabular}    
\end{center}
  
\begin{figure}[hbtp]
\centering
\includegraphics[scale=0.6]{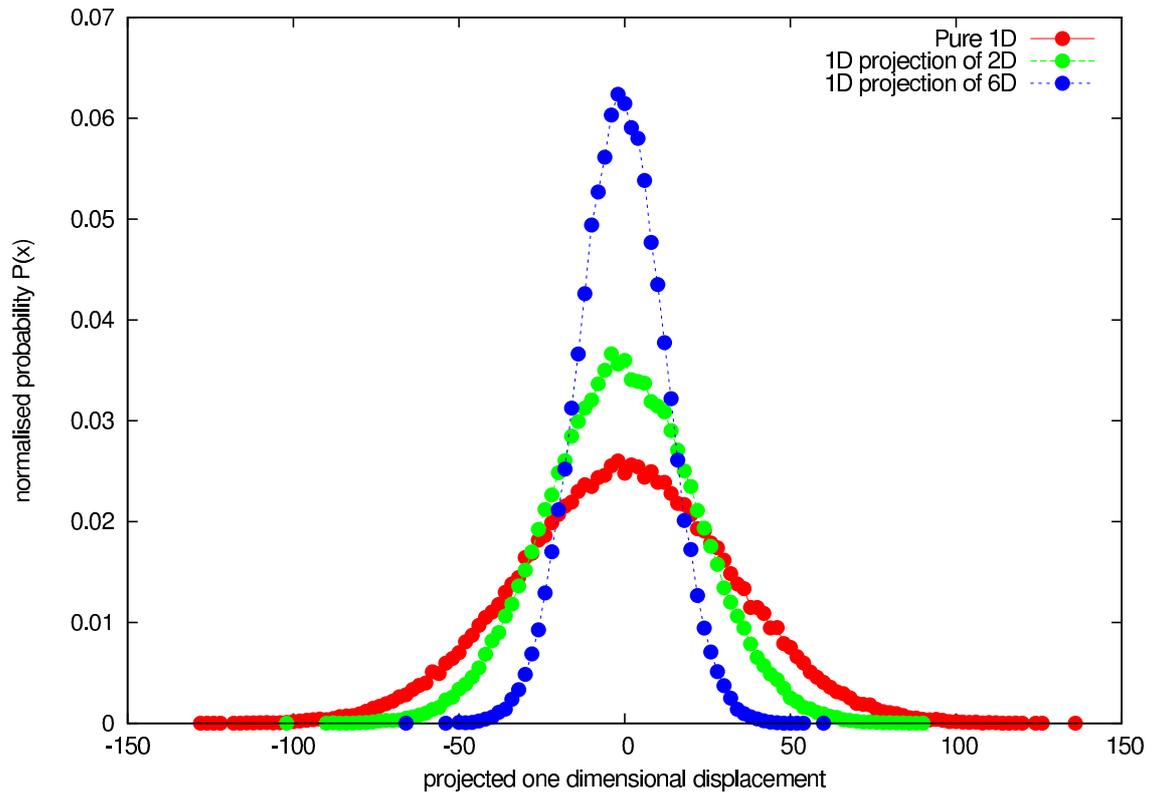}
\caption{Distribution of  displacement for 1 dimensional projection of 2 and 6 dimensions and pure one dimensional random walk  for $ N_{s} $=100000,$ N_{t} $=1000.}
\label{1d2d6ddistx}
\end{figure}

 \paragraph*{} 
 From the above analysis we get some similarities and dissimilarities between projected motion of generalised d-dimensional random walk with unit step length and pure 1 dimensional random walk with unit step length.:\subparagraph*{}(a) The distribution of   displacement for projected motion is Gaussian like pure 1 dimensional random walk with zero mean but the width of the distribution is not similar rather it is decreasing as the dimension is increasing (fig\ref{1d2d6ddistx}).\\(b) The mean-square displacement of projected motion ($\langle x^{2}\rangle$) is proportional to time i.e diffusive like pure 1 dimensional random walk but the diffusion-rate($ c $) is not same rather decreasing inversely as we increase the dimension. \paragraph*{}
 Now the question is, \textit{is there any generalized one dimensional random walk from which we can get similar statistical properties (diffusion rate,distribution of  displacements) of the projected motion of higher dimensional random walk with unit step length?} The answer can be given by the following analysis.

\begin{figure}[hbtp]
 \centering
 \includegraphics[scale=0.4]{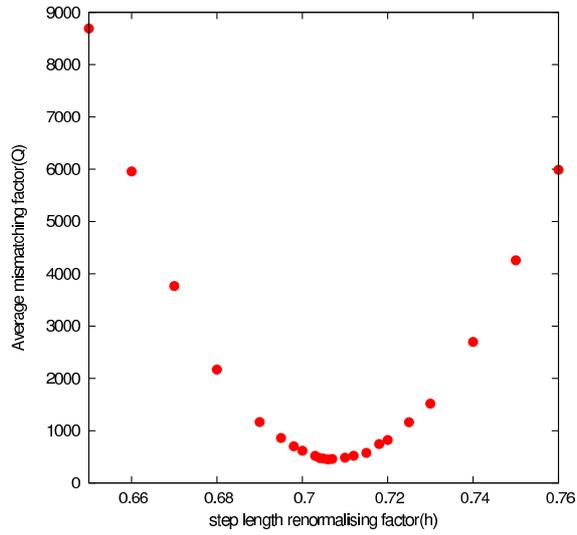}
 \caption{Average mismatching factor $(Q) $ versus step length renormalizing factor $(h)$ in 6 dimensions  for $ N_{s} $=100000,$ N_{t} $=1000}
 \label{mismatching6d}
 \end{figure}

\begin{figure}[hbtp]
\centering
\includegraphics[scale=0.4]{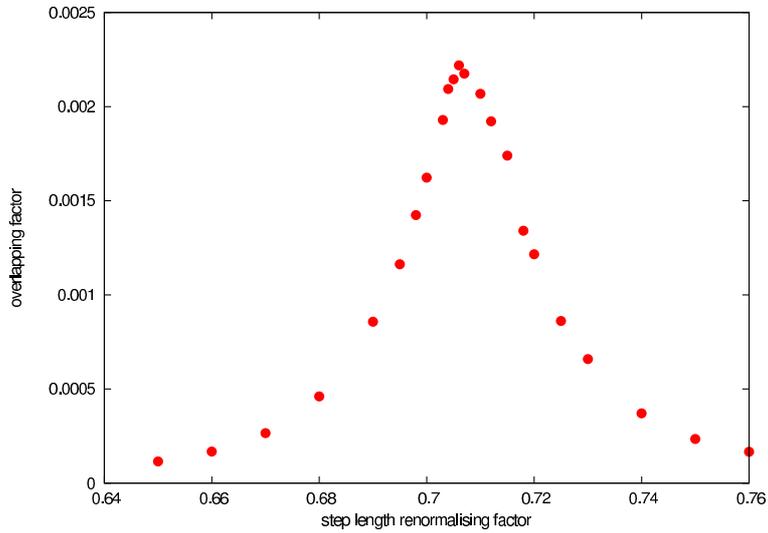}
\caption{Average overlapping factor (1/Q) (y axis in the graph) versus step length renormalizing factor (h) in 6 dimensions  for $ N_{s} $=100000,$ N_{t} $=1000. Here it is seen that the average overlapping factor is maximum for h=0.706 incase of 6 dimensions}
\label{overlap6d}
\end{figure}

\paragraph*{} From mathematical formalism (discussed above) it is clear that projected motion of higher dimensions (2, 3, 4,....d) is a random walk with random step length varying  between -1 to +1. But fig(5) shows that distribution of displacement of projected motion after $N_{t}=1000$ time steps are getting sharper and sharper as we increase the dimensions. So the possible reason for the distributions getting sharper and sharper may be the decrease of magnitude of projected step length $r_{1}$ as $d$ increases. From this consideration we can say that, the magnitude of $r_{1}$ changes along with dimensions and the boundary values of $r_{1}$ is no longer -1 to +1 rather it is different for different dimensions. So the ultimate question is how we can modify the pure one dimensional step length so that we can get equivalent statistical properties of projected motion from pure one dimensional random walk? In this regard we have introduced a factor $h$ named as \textit{step length renormalizing factor} and multiplied $r_{1}$ by this factor. Now $r_{1}$ varies between -h to +h instead of -1 to +1 and the value of h is different for different dimensions. For every dimension we choose h in such a way so that the average mismatching between two distributions (distribution of displacement after  $N_{t}$ times for projected motion and pure one dimensional motion) is minimum. We define the average mismatching factor by following way:\\Let, the probability density for pure 1 dimensional random walk is $P_{1D}(x)$ and probability density for projected motion is $P_{dDp}(x)$    
then the average mismatching factor $ Q$ is defined as 
\begin{equation}
Q=\frac{\sum_x(P_{dDp}(x)-P_{1D}(x))^2}{M}
\end{equation}
$ Q $ is obviously function of h i.e $Q=Q(h)$ and M is the number of points over which we are averaging. But estimating the value of h  from (in fig\ref{mismatching6d})is not precise as the minima is smeared. So we have plotted $ 1/Q(h) $ which is named as average overlapping factor, versus h where maxima is sharp and from this maxima we have estimated the value of h for every dimension precisely. For these values of $h$ corresponding to every dimension, the distributions(mentioned earlier) are almost similar i.e  overlap maximally(Fig\ref{overlap6d}).In this context it is to be mentioned that this definition of overlapping factor is not valid for $100$ percent accuracy(i.e if difference between $P_{dDp}(x)$  and $P_{1D}(x)$ is '0' , then $ 1/Q(h) $ goes to $\infty$).

\begin{figure}[hbtp]
\centering
\includegraphics[scale=0.6]{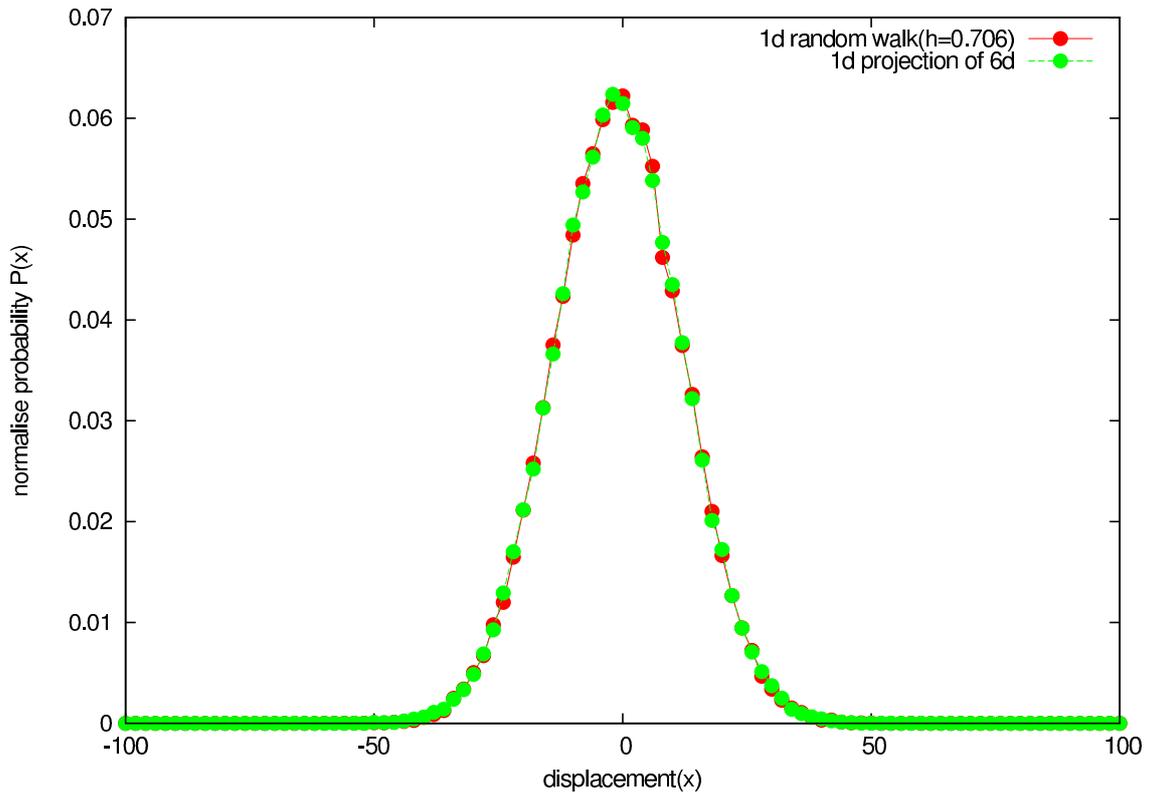}
\caption{Distribution of displacement of 1 dimensional projection of 6 dimensional random walk with unit step length and distribution of displacement of pure one dimensional random walk  with random step length varying uniformly between -h to h where h=0.706  for $ N_{s} $=100000,$ N_{t} $=1000. Here it is seen that the two distributions  overlap maximally for the estimated value of h .}
\label{1d6ddistx}

\end{figure}

\begin{figure}[hbtp]
\centering
\includegraphics[scale=0.6]{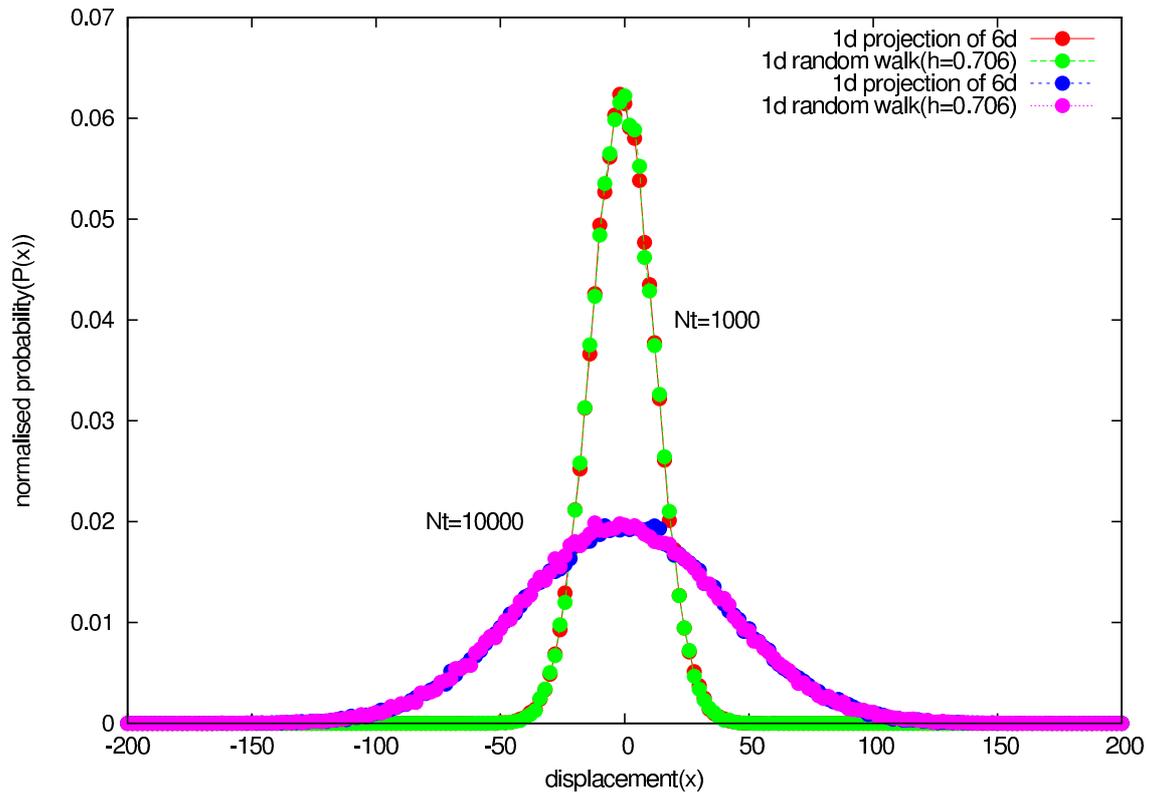}
\caption{Distribution of displacement of 1 dimensional projection of 6 dimensional random walk with unit step length and distribution of displacement of pure one dimensional random walk  with random step length varying uniformly between -h to h where h=0.706  for $ N_{s} $=100000,$ N_{t} $=1000 and $ N_{t} $=10000. Here it is seen that estimated value of h is invariant of $ N_{t} $.}
\label{1d6ddistxNt}
\end{figure}

\begin{figure}[hbtp]
\centering
\includegraphics[scale=0.6]{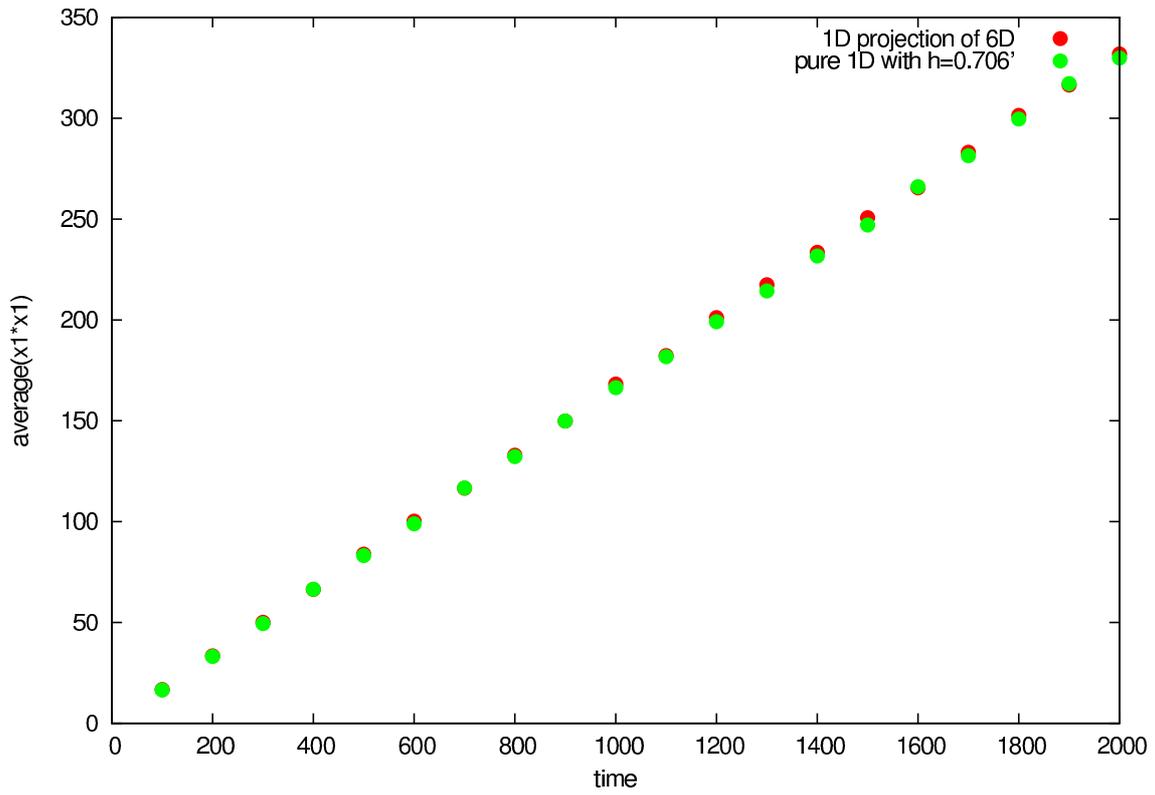}
\caption{Diffusion of 1 dimensional projection of 6 dimensional random walk with unit step length and pure 1 dimensional random walk with random step length varying uniformly between -h to h where h=0.706  for $ N_{s} $=100000.}
\label{diff1d6d}
\end{figure}

\paragraph*{}
Now In Fig\ref{1d6ddistx} it is seen that the distribution of displacement after time steps $ N_{t}$ for 1 dimensional projection of 6 dimensional random walk is almost similar with pure 1 dimensional random walk with random step length multiplying by h where h=0.706. Fig(\ref{1d6ddistxNt}) proves that estimated value of h is invariant of $N _{t} $. We have plotted $ \langle x^{2}\rangle $ versus $N _{t} $  which is almost similar. So the diffusive nature and the diffusion rate are also similar for these two motions (in fig\ref{diff1d6d}).\\Let us calculate the value of h in other dimensions in the same way.\begin{center}
In the following table numerical values (observed from simulation) of Step length renormalizing factor  corresponding to dimensions (2 to 10 dimensions) are mentioned:
\end{center}\begin{center}
\begin{tabular}{|c|c|}
\hline 
Dimensions & Step length renormalizing factor \\ 
\hline 

2 & 1.24 $ >1 $ \\
3 & 1.00 $ =1 $   \\
4 & 0.87$ <1 $ \\
5 & 0.77$ <1 $ \\
6 & 0.71$ <1 $ \\
7 & 0.65$ <1 $ \\
8 & 0.61$ <1 $ \\
9 & 0.58 $ <1 $\\
10 &0.55$ <1 $ \\
\hline 
\end{tabular}    
\end{center} Here in this table it is seen that for all dimensions h $ \leq $ 1 but in case of 2 dimensions it is 1.24 which is greater than 1(fig 30). It is a surprising result in the context that, as the walker moves with unit step length in two dimensions then its projected step length should be less than or equal to 1. But when we are modifying the pure one dimensional step length so that we can get equivalent statistical properties for projected motion from pure one dimensional motion then the multiplying factor h is greater than 1. In the following discussion we are trying to remove this confusion.\paragraph*{}
 In fig(\ref{distr1}) we have plotted the distribution of projected step length $r_{1}$ for different dimensions and pure one dimensional step length. Here it is observed that the step length of pure 1 dimensional random walk is uniformly distributed between (-h to h) (here it is plotted for h=1 but this is true for any value of h) but the  projected step length ($r_{1}$) are no longer uniformly distributed between -1 to 1 . As for example,for 2 dimensions, probability of getting large value of step length i.e near boundary (-1, +1) is high. So to have a comparison between  projected walk of 2 dimensions and pure 1 dimensional random walk , the magnitude of pure one dimensional step length should be increased . So the value of h is greater than 1. But for $d> 2$  probability of getting large value of step length (i.e near boundary) is low ; as we increase the dimension, the boundary value of $r_{1} $ decreases and the most probable value of $r_{1}$ is shifted towards 0. As a comprehensive effect the magnitude of $r_{1}$ is decreasing. Naturally the value of $h$ is also decreasing as dimension is increasing following the relation (fig\ref{hvsn}) \begin{equation}
  h\approx \dfrac{1.7}{\sqrt{d}} 
 \end{equation}
  It also supports the previous result eq(\ref{diff}) that for $d\rightarrow \infty$ there is no projected one dimensional  random walk.
 
 \paragraph*{}
  Now we are able to establish a generalized one dimensional random walk with which the  projected random walk can be compared. In all, it can be said that, we can get equivalent statistical properties of projected one dimensional motion of  d-dimensional random walk with unit step length from pure one dimensional random walk with random step length varying uniformly between -h to h where h is the step length renormalizing factor. In this context it is to be mentioned that in the simulation we have taken number of samples $N_{s}=100000$ and  $N_{t}=1000$. These choices of $N_{s}$ and  $N_{t}$ in this study are observed to be adequate to have stable statistics.

\newpage
\section{Concluding Remarks}
\paragraph{}In this paper we have discussed random walk problem extensively in generalised d-dimensional continuum by numerical simulation where we have observed how the statistical properties of random walker are changing with dimensions. Shifting of most probable value of the distribution of absolute displacement after a certain time ($N_{t}$) towards higher value with dimension is observed statistically. But the study of  projected one dimensional motion of higher dimensional random walk is our major area of interest where we have found that the width of the distribution of displacement of projected motion is getting sharper and sharper as we increase the dimensions. The projected motion is still diffusive irrespective  of any dimension, however the diffusion rate is changing inversely with dimensions.  As a consequence, we can predict that, \textit{\textbf{though in infinite dimensions there is a diffusion but its one dimensional projection is motionless}}. And finally we are able to make a good comparison between the projected one dimensional random walk and pure one dimensional random walk and came to the conclusion that\textit{\textbf{ one dimensional projection of generalised d-dimensional random walk with unit step length is equivalent to one dimensional pure random walk with random step length varying uniformly between $-h$ to $h$ where $h$ is the step length renormalizing factor.}}. Some more interesting studies can be done in this field like the first passage properties of random walker in higher dimensional continuum  on which we have already started to work.  Here we have reported only the numerical results of random walk in generalised d-dimensional continuum and its projected motion, but rigorous mathematical formulations of these are yet to be explored. And finally, there is a interesting aspect to think that, \textbf{\textit{why for getting equivalent statistical properties for the projected motion of three dimensional random walk from pure one dimensional random walk, the step length renormalising factor $h$ is exactly equal to 1 i.e the pure one dimensional step length is unrenormalised.}}

\newpage
\section{Acknowledgements}
We thank Amitava Banerjee and Abyaya Dhar for their help in preparing the figures and specially Amitava Banerjee for careful reading of the manuscript.

\newpage
\newpage
 \newpage
\newpage

\newpage 
\newpage
\newpage
\newpage
\newpage
\newpage
\newpage
\begin{figure}[hbtp]
\centering
\includegraphics[scale=0.6]{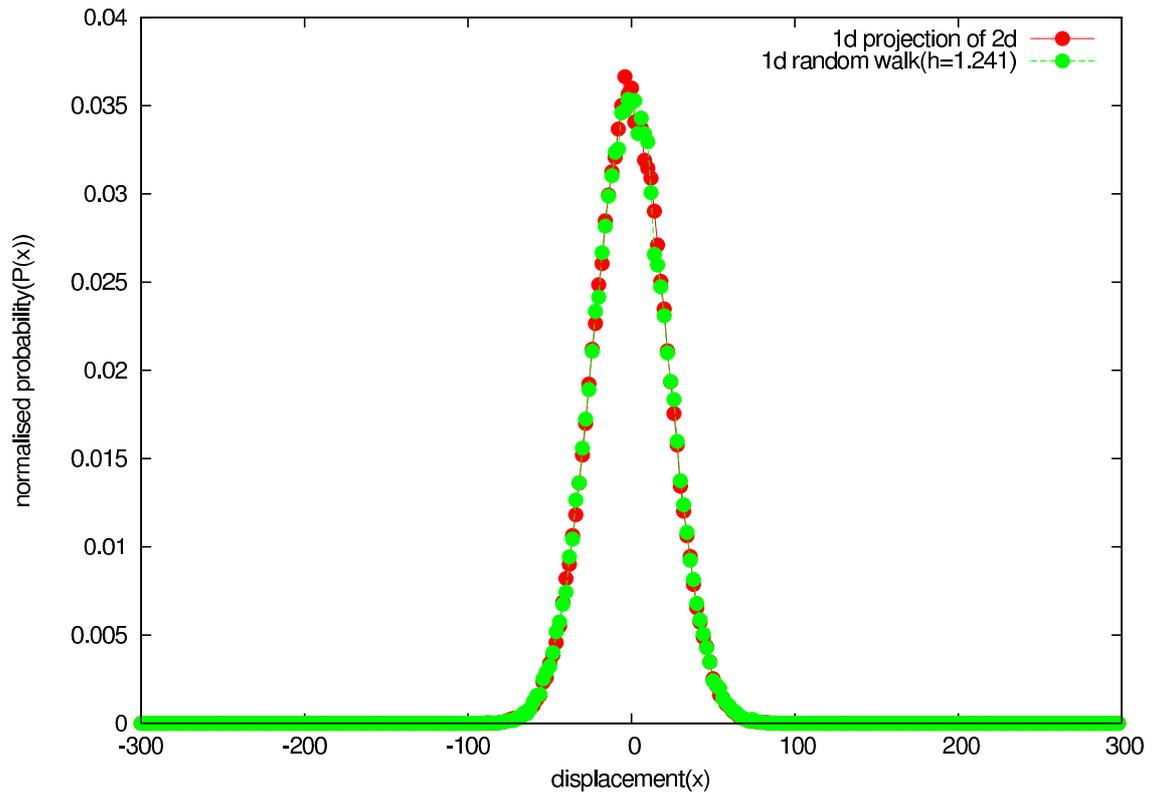}
\caption{Distribution of displacement of 1 dimensional projection of 2 dimensional random walk with unit step length and distribution of displacement of pure one dimensional random walk  with random step length varying uniformly between -h to h where h=1.241  for $ N_{s} $=100000,$ N_{t} $=1000. Here it is seen that the two distributions are almost same.}
\label{1d2ddistx}

\end{figure}
\newpage
\begin{figure}[hbtp]
\centering
\includegraphics[scale=0.6]{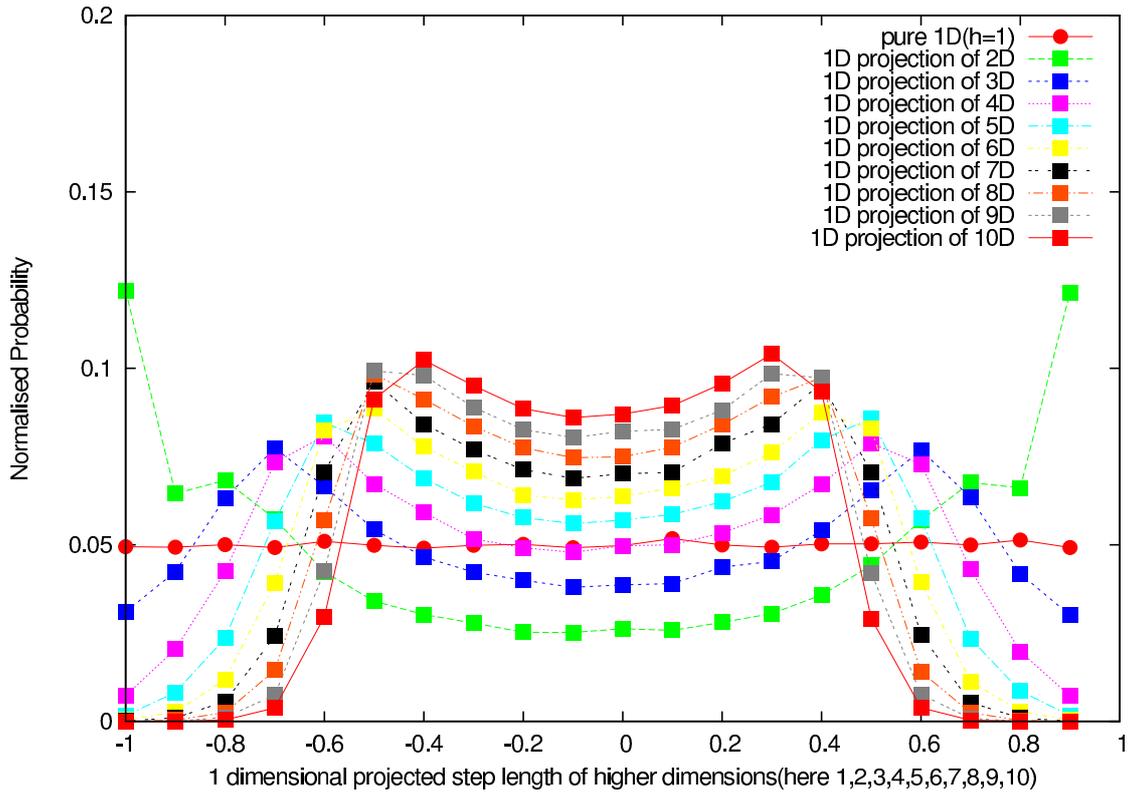}
\caption{Distribution of 1 dimensional projected step length of d-dimensional random walk (d= 2, 3, 4, 5, 6, 7, 8, 9, 10 dimensions ) and distribution of step length of pure 1 dimensional random walk where h=1 for $ N_{s} $=1000,$ N_{t} $=100.}
\label{distr1}
\end{figure}
\newpage

\begin{figure}[hbtp]
\centering
\includegraphics[scale=0.6]{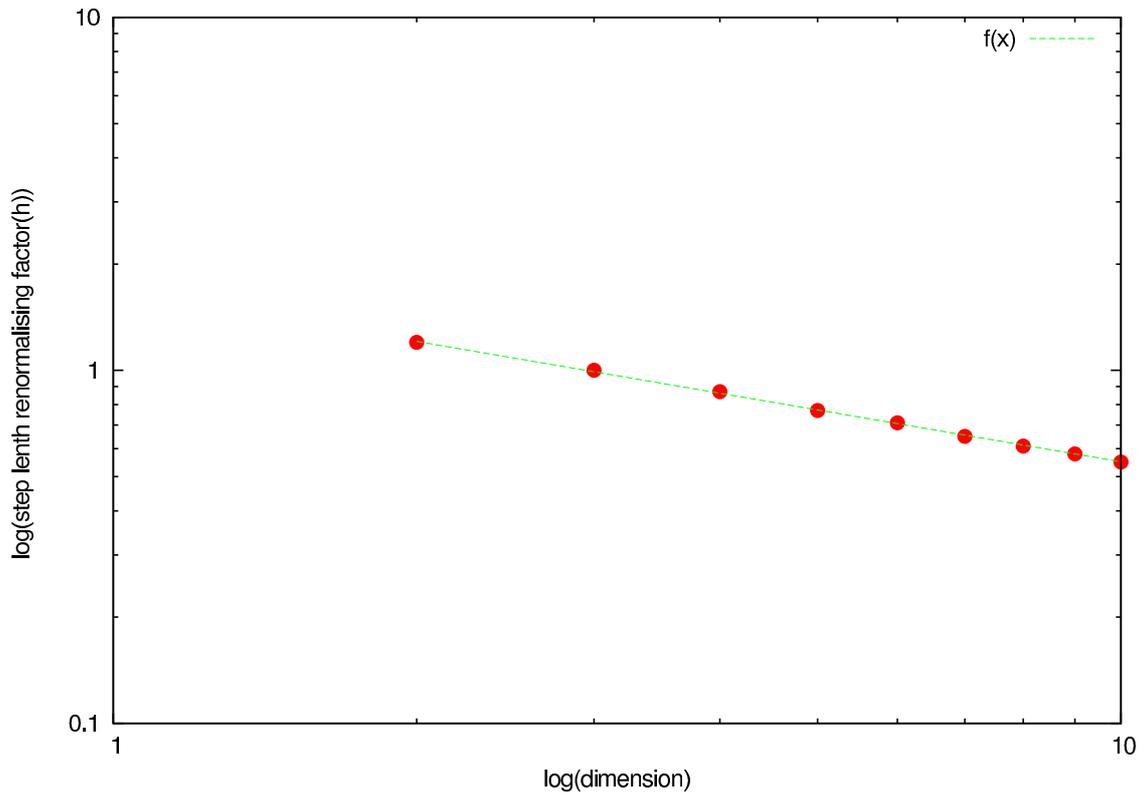}
\caption{ Step length renormalizing factor is plotted with different dimensions ($d$) (d= 2, 3, 4, 5, 6, 7, 8, 9, 10 dimensions ). The dotted line is $f(x)= 1.7/\sqrt{x})\ $ }
\label{hvsn}
\end{figure}

\end{document}